\documentclass{article}
\usepackage{frascatiphys,here,graphicx,subfigure}
\begin{document}
\title{ HEAVY QUARKONIA IN LIGHT-FRONT QCD }
\author{
Stanis{\l}aw D. G{\l}azek       \\
{\em Institute of Theoretical Physics, University of Warsaw, Poland } 
}
\maketitle
\baselineskip=11.6pt
\begin{abstract}
This talk is based on results obtained for masses
and wave functions of heavy quarkonia in a
light-front Hamiltonian formulation of QCD with
just one flavor of quarks using an ansatz for the
mass-gap for gluons. Since the calculated spectra
compare reasonably well with data, some further 
steps one can make are discussed.
\end{abstract}
\baselineskip=14pt
\section{Introduction}
Discussion of heavy quarkonium dynamics in this
talk is based on results for masses and wave
functions obtained in
Refs.\cite{Glazek,GlazekMlynik} in a relativistic
(boost-invariant) Hamiltonian formulation of QCD.
Steps involved in the calculation, starting with
the Lagrangian for QCD, deriving the corresponding
canonical light-front (LF) Hamiltonian, carrying
out the renormalization group procedure for
effective particles (RGPEP) to obtain the quark
and gluon operators at finite momentum scales,
$\lambda$, deriving an effective Hamiltonian for
heavy constituent quarks and gluons, $H_\lambda$,
at momentum scales on the order of the quark mass,
$\lambda_0 \sim m$, using an ansatz for the
mass-gap for gluons, $\mu$, to finesse a new
Hamiltonian that acts only in the effective
quark-antiquark Fock sector, $H_{\lambda_0 Q\bar
Q}$, and solving numerically the resulting
eigenvalue problem for $H_{\lambda_0 Q\bar Q}$,
are described in the original literature (a
condensed summary\cite{QCD2006} is available).

Here, only one example of results for quarkonium
masses obtained from $H_{\lambda_0 Q\bar Q}$ is
quoted, to illustrate what happens in the simplest
version of the LF approach to QCD. The key point
is that the results do not depend on the ansatz
$\mu$ and fit data reasonably well for the
coupling constant expected from RGPEP, assuming it
has a known value at $\lambda = M_Z$, and for the
charm or bottom quark masses that have typically
considered sizes. Then, the emerging recipe for
the mass gap ansatz as a tool to facilitate
numerical studies of effective quark and gluon
dynamics is described. The ansatz is designed to
be introduced only in the final stage of
diagonalizing $H_{\lambda_0 Q\bar Q}$.

LF quantum field theory has a long
history\cite{BrodskyPauliPinsky} with lots of
modern developments\cite{ILCAC} that cannot be
duly reviewed here. As part of the progress, a
conceptual outline of nonperturbative QCD in the
LF frame was achieved\cite{NPQCD} using the
similarity renormalization group
procedure\cite{similarity1,similarity2}. A
confining logarithmic potential of order $\alpha$
in quark-antiquark sector has been
discovered\cite{PerryBrazil} and studied in heavy
quarkonia\cite{Brisudova1,Brisudova2} and other
systems\cite{Allen,Kylin2}, but not using
RGPEP\cite{Brisudova4}. Besides the work that led
to Refs.\cite{Glazek,GlazekMlynik}, distinct RGPEP
applications in scattering theory\cite{Wieckowski}
and gluonium\cite{Maslowski} are relatively new.
But there is a lot of work to do in LF QCD before
it can be widely accepted as a viable alternative
to lattice QCD\cite{WilsonLattice}. The AdS/CFT
method\cite{AdSCFT1,AdSCFT2} has not been
connected with RGPEP in QCD yet.
\section{ Bottomonium masses as example }
In the crudest version, $H_\lambda$ is calculated
using RGPEP to first order in $\alpha_{\lambda_0}$
in one flavor QCD, and $H_{\lambda_0 Q\bar Q}$ is
then evaluated to the same order, using the ansatz
$\mu$ for gluons in the effective
quark-antiquark-gluon sector (see next section).
The resulting eigenvalue problem does not depend
on the ansatz $\mu$ and takes a form that satisfies
requirements of rotational symmetry despite that
the LF reference frame distinguishes $z$-axis. For
example, the eigenstates with quantum numbers of
$\Upsilon$ are described by a $2 \times 2$ matrix
wave function $\phi(\vec k \,) = \vec b (\vec k\,)
\vec \sigma$, where ($\vec s$ is the polarization
three-vector that determines the polarization
state of the whole quarkonium in motion with
arbitrary velocity)
\begin{eqnarray}
b^m(\vec k \,)
 = 
\left[ \delta^{mn} \, { S(k) \over k} + 
{1 \over \sqrt{2}} \, 
\left( \delta^{mn} - 3  { k^m k^n \over k^2 } \right) \, 
{ D(k)  \over k } \right] \, s^n \, . 
\end{eqnarray}
The matrix $\phi$ enters into the definition
of a relativistic quantum state of a quarkonium 
with definite momentum and mass $M$,
\begin{eqnarray}
\label{state}
|M,P^+,P^\perp, \vec s \, \rangle 
& = & \int [ij] \, 
(2\pi)^3 P^+ \delta^3(P - p_i - p_j) \,
\chi^\dagger_i \, \phi( \vec k_{ij}) \chi_j \,
b^\dagger_{\lambda_0 i} d^\dagger_{\lambda_0 j} \,
|0\rangle  \, , \nonumber \\
& & 
\end{eqnarray}
where $i$ and $j$ denote flavor, momentum, and
spin quantum numbers (colors are combined
to 0) of scale-dependent effective quarks that are
created by operators $b^\dagger_{\lambda_0 i}$ and
$d^\dagger_{\lambda_0 j}$ from the LF QCD vacuum.
These LF operators depend on the scale $\lambda$
(in ratio to $\Lambda_{QCD}$ in the RGPEP scheme,
and the quark mass). The relative momentum
three-vector $\vec k_{ij}$ is defined using LF
kinetic momentum variables and the quark mass
corresponding to the scale $\lambda_0$. The
eigenvalue equation satisfied by the $S$ and $D$
wave functions is written in terms of
dimensionless momentum variables
\begin{eqnarray}
\vec p = \vec k_{ij} /k_B \, ,
\end{eqnarray}
where $k_B$ denotes the strong Bohr momentum,
$\alpha_{\lambda_0} m_{\lambda_0}/2$ (subscript 
$\lambda_0$ will be dropped from now on). Using 
$p = |\vec p \,|$, the radial equation can be
written as
\begin{eqnarray}
\label{Upsilon}
\left[
\begin{array}{cc}
h_{osc}  & 0  \\
0  & h_{osc} + k_p {6 \over p^2}
\end{array}
\right]
\left[
\begin{array}{c}
S(p)  \\
D(p)
\end{array}
\right]
 = 
\int_0^\infty \hspace{-3mm} dk \, f {2 p k \over \pi} 
\left[
\begin{array}{cc}
{\cal W}_{ss} &  {\cal W}_{sd} \\
{\cal W}_{ds} &  {\cal W}_{dd}
\end{array}
\right]
\left[
\begin{array}{c}
S(k)  \\
D(k)
\end{array} 
\right] \nonumber \\
\end{eqnarray}
with
\begin{eqnarray}
h_{osc} 
& = & 
p^2 - k_p \partial_p^2 - x \, , \\
k_p 
& = &
{9 \over 128 \, \sqrt{2 \pi}} \, \left( {
\lambda_0^2 \over \alpha \, m^2 } \right)^3 \, ,
\end{eqnarray}
while the quarkonium mass eigenvalue is given by the 
eigenvalue $x$ through 
\begin{eqnarray}
M = 2m \, \sqrt{ 1 + x \, \left( {2\over 3} \alpha
\right)^2 }  \, .
\end{eqnarray}
Note that the eigenvalue is not energy in any
specific frame of reference but the mass itself.
The functions ${\cal W}_{ss}$, ${\cal W}_{sd}$, 
${\cal W}_{ds}$, ${\cal W}_{dd}$ are given in the 
literature\cite{GlazekMlynik}. 

There is no quantitative trace of the gluon mass
ansatz in this result. But there is a
qualitatively new element in the form of a
harmonic oscillator correction to the strong
Coulomb potential (with LF Breit-Fermi terms). 

Another qualitatively new element, a result of 
using RGPEP, is the form factor 
\begin{eqnarray}
f & = & 
\exp{\left\{ - \left[ { {\cal M}^2(p) - {\cal M}^2(k) \over \lambda_0^2 } 
\right]^2 \right\} } \, , 
\end{eqnarray}
where $\cal M$ denotes an invariant mass of a pair
of free quarks. The form factor tempers the
spin-dependent gluon exchange interaction. In
particular, it regulates otherwise
ultraviolet-divergent three-dimensional delta
functions (in the position space formally 
associated with the momentum space of $\vec k$
via the Fourier transform), which are present in
the functions $\cal W$ due to the relativistic 
spin effects. 

One solves the eigenvalue equations for $b \bar b$
bound states, such as eq.\ref{Upsilon}, assuming
that $\alpha$ is given by the RGPEP evolution from
the known value at $\lambda = M_Z$ down to
$\lambda_0$. If $\alpha_{M_Z} = 0.12$, the lowest
order RGPEP evolution of $\alpha_\lambda$ in QCD
with only one flavor\cite{glambda} produces
$\alpha \sim 0.326$ at $\lambda_0 \sim 3.7$ GeV
(about 30\% smaller value is generated for 6 or
5 flavors). Less is known about the RGPEP evolution
and value of the $b$-quark mass, $m_b$.
Tab.\ref{Upsilontab} shows masses of $b \bar b$
quarkonia obtained\cite{GlazekMlynik} when
$\alpha$ and $m_b$ are adjusted to reproduce
masses of $\chi_1$(1P) and $\chi_1$(2P) at
$\lambda_0 = 3697.67$ MeV. 

\begin{table}[htb]
\centering
\caption{ Example of calculated masses (MeV) for $b \bar b$
states. The corresponding coupling constant and quark mass
are $\alpha$ = 0.32595 and $m_b$ = 4856.92 MeV.} 
\vskip 0.1 in
\begin{tabular}{|l|l|l|r|} 
\hline
meson           &     theory    & experiment\cite{PDG}        & difference \\
\hline
\hline
$\Upsilon$10860 &     10725     &  \parbox{1.4cm}{10865   }$\pm$8             &   -140  \\
$\Upsilon$10580 &     10464     &  \parbox{1.4cm}{10579.4 }$\pm$1.2           &   -116  \\
$\Upsilon$3S    &     10382     &  \parbox{1.4cm}{10355.2 }$\pm$0.5           &     27  \\
$\chi_2$2P      &     10276     &  \parbox{1.4cm}{10268.65}$\pm$0.22$\pm$0.50 &      7  \\
$\chi_1$2P      &     10256     &  \parbox{1.4cm}{10255.46}$\pm$0.22$\pm$0.50 &      0  \\
$\chi_0$2P      &     10226     &  \parbox{1.4cm}{10232.5 }$\pm$0.4 $\pm$0.5  &     -6  \\
$\Upsilon$2S    &     10012     &  \parbox{1.4cm}{10023.26}$\pm$0.31          &    -11  \\
$\chi_2$1P      &     ~9912     &  \parbox{1.4cm}{~9912.21}$\pm$0.26$\pm$0.31 &     -1  \\
$\chi_1$1P      &     ~9893     &  \parbox{1.4cm}{~9892.78}$\pm$0.26$\pm$0.31 &      0  \\
$\chi_0$1P      &     ~9865     &  \parbox{1.4cm}{~9859.44}$\pm$0.42$\pm$0.31 &      5  \\
$\Upsilon$1S    &     ~9551     &  \parbox{1.4cm}{~9460.30}$\pm$0.26          &     91  \\
$\eta_b$1S      &     ~9510     &  \parbox{1.4cm}{~9300   }$\pm$20  $\pm$20   &    210  \\
\hline
\end{tabular}
\label{Upsilontab}
\end{table}

If the RGPEP calculation of $H_\lambda$ and
subsequent reduction to $H_{\lambda Q \bar Q}$
were exact, there should be no dependence of the
spectrum on $\lambda$. Once $\alpha$ and $m_b$ are
adjusted to observables at one scale, they evolve
in some exact way, including the formation of
bound states. But in this crudely simplified
version of LF QCD, the RGPEP procedure is limited
to order $\alpha$ and $H_{\lambda Q \bar Q}$ is
finessed using an ansatz for the gluon mass gap
$\mu$. Therefore, one cannot change $\lambda$
considerably using equations limited to order
$\alpha$ and there is only a hope that in
some small range of values of $\lambda_0$ the
equations have a chance to work once the coupling
constant and quark mass are given their right
relativistic values\cite{NPQCD,GlazekWilson}.

There is a characteristic pattern visible in the
fourth column in tab.\ref{Upsilontab}: the greater
the difference between a mass eigenvalue and the
masses of quarkonia in the middle of the table,
used to choose $\alpha$ and $m_b$, the greater the
discrepancy between the crudely approximated
theory and experiment. This should be expected.
The most strongly bound states are sensitive to
deviations of the effective potential from the
Coulomb shape. For example, interactions order
$g^4$ (or $\alpha^2$), introduce
$\delta$-functions that are absent here because of
the limitation to terms order $g^2$ in the RGPEP
and two-quark reduction. Analogous 4th order
$\delta$s and other singular corrections are known
in QED. Here such terms should have much larger
effect because the coupling constant is about 30
times larger then in QED (they have to be treated
nonperturbatively). The least strongly bound
states, those with largest masses, should not be
described well without proper inclusion of gluons.
The mass ansatz should fail to render interactions
that are associated with gluons producing a linear
potential at distances much larger than the strong 
Bohr radius. 
\section{ Mass ansatz as a computational tool in QCD }

The mass ansatz for virtual effective particles in
Fock components that contain more such particles
than just two in mesons, or three in baryons,
deserves a comment for several reasons. One reason
is that the ansatz may help improve the
calculations for heavy quarkonia. Another reason
is that it may lead to a possibility of
calculating properties of baryons built from heavy
quarks. The third reason is that it may help in
crossing the barrier that separates all
small-coupling expansions in QCD from entering the
region of quark masses much smaller than
$\Lambda_{QCD}$. 

The first two reasons concern the difficulty that
precise numerical solutions of eigenvalue problems
with coupled two-, three-, and more-particle
sectors are hard to obtain. In this respect, the
mass ansatz appears a candidate to mimic what
happens in an infinite tower of the Fock
components built from effective particles. Since
the RGPEP form factors limit momentum transfers by
$\lambda$, the spread of probability to sectors
with many effective particles corresponding to the
scale $\lambda$ is tamed. But these sectors do
influence the dynamics of the dominant sectors and
some ansatz appears inevitable. The question is
how to make it self-consistently. The basic idea
is to drop all sectors above the highest included
(in the sense of number of the effective
particles), put in instead a mass ansatz in the
highest Fock component, and see what happens in
the dynamics of lower components. The next step is
to increase the maximal number of effective
particles by one and see if the same type of
ansatz is producing the same answers. A small
coupling expansion may constrain the options
sufficiently for finding good candidates for
suitable mass terms in the highest components.

The third reason that concerns light quarks is most
speculative. It involves chiral symmetry, or
rather the mechanism of its breaking. In LF
Hamiltonian of QCD with small $\lambda$, there may
exist finite terms that violate chiral symmetry
and do not vanish in the limit of quark mass
approaching zero\cite{NPQCD}. At the same time,
the LF vacuum state remains simple due to cutoffs
imposed on the particle momentum components along 
the front. The question is how to find those terms 
in practice. LF power counting limits the structure 
of allowed terms but so far insufficiently for 
anybody to tackle the issue, even though the stakes 
are high. 

From this point of view, the following observation
is of interest. Consider a colorless state built
from two effective gluons. They attract each
other. Consider then that one of these gluons turns
into a pair of quarks. These quarks are in an
octet state and instead of attracting they repel
each other. In perturbation theory, if the number
of quark flavors is not too high, this is not
dangerous and gluonic interactions sustain
asymptotic freedom, generating infrared slavery.
However, beyond perturbation theory, a pointlike
creation of a pair of quarks that repel each other
by violent potentials may lead to an explosive
behavior. Such behavior is entirely absent in QED
because electrons attract anti-electrons and this
effect slows down the growth of a pair, instead of
accelerating it. 

To be more specific, consider eq.\ref{Upsilon}, in
which the $s$-wave potential ${\cal W}_{ss}$
contains a term\cite{GlazekMlynik} that would be a
$\delta$-function if the form factor $f$ did not
smooth it out. This term is attractive in
$\Upsilon$. It is repulsive instead in the color
octet states. When $\lambda$ is comparable to the
heavy quark mass, which means that $\alpha$ is
small, the smallness of the coupling constant and
form factor width produce an interaction that
cannot compete significantly with the size of the
quark mass. This is visible in tab.\ref{Upsilontab}. 
However, if the quarks are light, it is entirely 
unclear what will happen.

The situation is different than in the case of 
analogy between a gluonium and a helium atom 
with one doubly charged electron discussed in 
Ref.\cite{GlazekWilson}. Here, two particles 
with the same charge are suddenly put on top of 
each other and large potential energy is created. 

In the case of light quarks, the large terms are
smoothed out by the RGPEP form factors $f$ but
their strength may be comparable with
$\Lambda_{QCD}$. The central point is that in
order to find out what happens due the explosive
nature of color dynamics of effective particles
beyond perturbation theory, one has to separate
some sectors from a presumably decreasing but in
principle infinite chain of them. This is what the
mass ansatz facilitates. Thus, it opens a way to
investigate interactions in the effective
Hamiltonians that may be calculated perturbatively
in RGPEP and then diagonalized nonperturbatively
using computers. This way one can find out if the
effective interactions may in principle be
responsible for emergence of the constituent quark
masses for quarks $u$, $d$, and $s$.

\section{ Conclusion } 
Tab.\ref{Upsilontab} shows that the
chance that LF Hamiltonian approach to QCD may
apply in phenomenology of heavy quarkonia is not
hopelessly small. Since the coupling constant one
needs is order 1/3, and one may need even a
smaller coupling constant when 4th order RGPEP is
used, the approach stands ready for a more
extended scrutiny. The reason it deserves to be
checked is that it appears now to indicate a
possibility that a single formulation of the
entire theory with quark masses much greater than
$\Lambda_{QCD}$ is conceivable with no need to
combine different formulations for including
information concerning different scales. 

The harmonic potential finessed in the
quark-antiquark sector using an ansatz for the
gluon mass gap, leads to the eigenvalues $M^2$
that are proportional to the angular momentum of
relative motion of quarks, like in the Regge
trajectories. It is found that the oscillator
frequencies are on the order of one inverse fermi,
and the oscillator potential grows as the relative
distance squared in fermis with a coefficient
given by the quark mass. So, presumably, for
states with masses greater than order 1 GeV above
the ground states, the probability of emission of
effective gluons increases and then formation of
strings of gluons is favored if the gluons also
have some oscillator force acting among them. For
quantum gluons to form a string, each pair of the
neighboring gluons must be held together stronger
than by a linear potential. If it is capable of
generating such effects, the LF Hamiltonian
approach could thus lead to a quantum theory of
the gluon string in QCD without ever introducing a
nontrivial vacuum. 

But the concept of mass ansatz in Fock sectors
with one more effective particle than the maximal
number treated nonperturbatively, is probably most
interesting as a tool for finding out what happens
when one attempts to solve eigenvalue equations
for Hamiltonians that are evaluated using
perturbative RGPEP in the case of canonical QCD
with quark masses smaller than $\Lambda_{QCD}$.
The idea discussed here is that effective
particles in non-singlet color configurations may
experience explosive potentials in the form of
smoothed $\delta$-functions that may cause
effects order $\Lambda_{QCD}$ per constituent. The
ansatz for quark and gluon masses inserted in the 
highest sectors thus opens a possibility to generate 
concrete forms of such terms and to study them  
nonperturbatively. Nothing is known yet about 
what may come out from such studies.
\section{Acknowledgements}
I would like to thank Robert Perry for comments on the
manuscript and acknowledge the outstanding hospitality
of the organizers.
\end{document}